\documentclass[12pt,letterpaper]{article}
\usepackage{setspace}
\usepackage[margin=1.1in]{geometry}

\usepackage{epsfig, amsthm, amsmath, amssymb, latexsym, xpatch, xcolor}
\usepackage[titletoc]{appendix}
\usepackage{float} 
\usepackage[margin=0.5cm, font={small,it}]{caption}
\usepackage{wrapfig}
\usepackage[normalem]{ulem} 

\usepackage[mathscr]{euscript} 

\newcommand{\oop}{\begin{quote}}
\newcommand{\eed}{\end{quote}}

\usepackage{graphicx}

\theoremstyle{plain}
\newtheorem*{theorem*}{Theorem}

\newtheorem*{lemma*}{Lemma}

\theoremstyle{definition}

\newtheorem*{definition*}{Definition}

\newtheorem*{example*}{Example}

\newcommand{\op}{\begin{itemize}}
\newcommand{\ed}{\end{itemize}}

\newcommand{\opp}{\begin{quote}}
\newcommand{\edd}{\end{quote}}

\newcommand{\ope}{\begin{enumerate}}
\newcommand{\ede}{\end{enumerate}}

\newcommand{\im}{\item}

\newcommand{\PP}{\mathsf{Pr}}

\newcommand{\given}{\mbox{$\,|\,$}}

\title{The Problem of the Priors, or Posteriors?}

\author{Hanti Lin \\[0.5em] University of California, Davis \\ika@ucdavis.edu}

\date{} 

\begin{document}

\maketitle

\begin{abstract} \noindent The problem of the priors is well known: it concerns the challenge of identifying norms that govern one's prior credences. I argue that a key to addressing this problem lies in considering what I call the problem of the posteriors---the challenge of identifying norms that directly govern one's posterior credences, which backward induce some norms on the priors via the diachronic requirement of conditionalization. This forward-looking approach can be summarized as: {\em Think ahead, work backward}. Although this idea can be traced to Freedman (1963), Carnap (1963), and Shimony (1970), I believe that it has not received enough attention. In this paper, I initiate a systematic defense of forward-looking Bayesianism, addressing potential objections from more traditional views (both subjectivist and objectivist). I also develop a specific approach to forward-looking Bayesianism---one that values the convergence of posterior credences to the truth, and treats it as a fundamental rather than derived norm. This approach, called {\em convergentist Bayesianism}, is argued to be crucial for a Bayesian foundation of Ockham's razor in statistics and machine learning. 
\\[0.6em] Keywords: {\em Bayesian Epistemology}, {\em Convergence to the Truth}, {\em Ockham's Razor}, {\em Statistics}, {\em Machine Learning}
\end{abstract}

\tableofcontents

\section{Introduction}

The problem of the priors in Bayesian epistemology is the challenge of identifying the norms that govern one's prior credences---the credences at the start of one's inquiry. Two traditional approaches are well known: {\em subjective Bayesians} think that priors are only required to be coherent (pending an account of coherence); objective Bayesians think that one's distribution of prior credences should be not just coherent but also as flat as possible (pending an account of flatness). Those two traditional approaches share something in common: they both only impose normative constraints directly on priors. This paper defends an alternative approach: in addition to norms that directly regulate the priors, let's also consider another type of norm, {\em forward-looking} norms, which impose a constraint directly on posterior credences. An early example is Shimony's (1970) norm of Open-Mindedness, which requires (roughly) that each seriously considered hypothesis have the possibility of receiving a high posterior credence. 

How might norms on posteriors be relevant to the problem of the priors? The answer can be illustrated with a seesaw. Note that there is a joint constraint on the two sides of a seesaw. So, if we restrict the motion of any one of the two sides, it leads to a constraint on the other side. Now, the norm of Conditionalization is like a seesaw, placing a joint constraint on the ``prior'' and ``posterior'' sides. While the traditional approaches impose a constraint directly on the ``prior'' side of the Conditionalization seesaw in order to generate a constraint on the ``posterior'' side, I propose to also consider the opposite direction. 

In general, a forward-looking norm is a norm that imposes a constraint directly on posterior credences, which might be able to  {\em backward induce} a constraint on prior credences---under the assumption of Conditionalization as a joint constraint on both priors and posteriors. So, a forward-looking norm can thus regulate priors, albeit only indirectly---through the mediation of Conditionalization. The idea is straightforward: {\em Think ahead, work backward.} I call this idea {\em forward-looking Bayesianism}, a somewhat heretical approach to the problem of the priors. 

Although forward-looking Bayesianism is rarely, if ever, identified as a distinctive approach to the problem of the priors, its idea has a long history, dating back to the 1960s. The statistician Freedman (1963) proposed a forward-looking norm on posterior credences---specifically, a convergence norm requiring a certain type of convergence of posteriors to the truth.
	That same year, the philosopher Carnap (1963) proposed a different convergence norm on posteriors to justify (or rationally reconstruct) a version of enumerative induction.\footnote{That norm is called ``A13. Axiom of Convergence'' by Carnap (1963, p. 976), who incorporates some ideas from Reichenbach's (1938) convergentist, non-Bayesian epistemology.}
Soon after, Shimony (1970) advocated the norm of Open-Mindedness as mentioned above. 

Despite its long history, forward-looking Bayesianism initially gained few followers in philosophy and---perhaps more tellingly---drew little criticism; the prevailing response was one of indifference. Fortunately, the forward-looking approach has recently gained traction, pursued by Belot (2013),\footnote
	{
	Belot (2013) advocates a forward-looking norm---a variant of Shimony's Open-Mindedness---and uses it to challenge a traditional Bayesian view. This has prompted many responses defending the more traditional perspective (Huttegger 2015, Weatherson 2015, Elga 2015, Cisewski et al. 2018). Interestingly, however, these critiques target Belot's other assumptions while being largely OK with his Open-Mindedness norm---a welcome step, in my view, toward embracing forward-looking Bayesianism.
	} 
Huttegger (2017),\footnote
	{
	Huttegger (2017, p.\,170) outlines how a forward-looking norm---specifically, a convergence norm requiring posteriors to converge to the truth---may be used to support a controversial constraint on priors: Countable Additivity. More on this in Section \ref{sec-countable}.
	}
and Lin (2022).\footnote
	{
	Lin (2022) first develops a convergentist epistemology for beliefs, but then extends it to credences, with a convergence norm that induces backward a normative constraint on priors, aiming to justify a version of enumerative induction and Ockham's razor in the face of severe underdetermination of theory by data.
	}

These recent contributions, however, only seek to promote {\em specific} norms on posteriors rather than defending the {\em general} framework of forward-looking Bayesianism, under which the followers may disagree on the correct norms on posteriors. So, here, I will articulate the core thesis of the framework (Section \ref{sec-core}), trying to unite proponents and form a faction that can compete with the two traditional ones, subjectivist and objectivist. This task is important to help me identify and address some general worries about the framework itself rather than any particular norm on posteriors. 
An example of such a general worry is that, once an agent has probabilistic prior credences, the norm of Conditionalization will do the rest of the work to completely determine what posterior credences ought to be like:
	\opp 
	{\bf Conditionalization}. The following ought to be the case whenever the denominator is nonzero, where $E$ denotes the available evidence:
	\begin{eqnarray*}
	\text{Posterior credence in } H
	&=& \text{Conditional prior credence in } H \text{ given } E 
	\\[0.5em]
	&=& \frac{\text{Prior credence in } (H \wedge E)}{\text{Prior credence in } E} \,.
	\end{eqnarray*}
	\eed 
So, there appears to be no need for any additional constraints on the posteriors---no epistemic role for forward-looking norms. I will respond to this objection below (Sections \ref{sec-default-bayes} and \ref{sec-ought-to-be}). 

Once forward-looking Bayesianism is defended as a general framework, the remainder of the paper will develop my preferred implementation of it---{\em convergentist} Bayesianism. Specifically, I will defend some convergence norms, which require posterior credences to converge to the truth in a context-sensitive manner, where the truth is simply the true answer to the question one pursues in one's context of inquiry. As you might expect, I will address the all-too-familiar Keynesian worry that convergence is epistemically irrelevant, given that we are all dead in the long run (Section \ref{sec-death-long}).

In addition to addressing objections, I will offer a positive argument for convergentist Bayesianism---and thereby for forward-looking Bayesianism more broadly. If I am correct, the convergentist approach holds the key to justifying Ockham's razor in scientifically important cases, such as the problem of curve fitting that comes {\em without} assuming the truth of any particular parametric model, whether the linear model $Y = a X + b$, the quadratic model $Y = a X^2 + b X + c$, or the like. I will argue that such a (nonparametric) problem of curve fitting can be adequately addressed only by convergentist Bayesianism, not by the more traditional views---subjectivist or objectivist. If so, Bayesians' best hope for a foundation of machine learning and artificial intelligence lies in the convergentist approach, since machine learning research generally seeks to do without parametric assumptions. The ideas will be detailed below (Section \ref{sec-nonparametric}).

Before delivering on the promises I have made, let me start by articulating the view I aim to defend.

\section{What Is Forward-Looking Bayes?}\label{sec-core}

I propose defining forward-looking Bayesians as those who endorse the following:
\oop 
{\sc The Core Theses of Forward-Looking Bayesianism}. 
\ope 
\im {\bf Thinking Ahead}. Some norms, called forward-looking norms, directly govern the posterior credences that an agent has in response to evidence. 

\im {\bf Working Backward}. Some forward-looking norms can be used to backward induce substantive constraints on prior credences, via the diachronic norm of Conditionalization. 

\im {\bf Fundamentality}. Some forward-looking norms are as fundamental as the traditional package, which includes Conditionalization and some synchronic norms on priors such as Probabilism. Those forward-looking norms promote some epistemic virtues that the traditional package fails to accommodate.
\ede 
\eed 
\noindent The ideas underlying this view will be illustrated with examples and compared to the more traditional views in this section. This will clarify the view before I defend it in the sections that follow.

\subsection{Preliminaries}\label{sec-raven}

For concreteness, I will often ground my discussions on a simple empirical problem. Imagine a scientist pursuing the following empirical problem: 
	\oop 
	{\it Example: The Raven Problem}.
	\op 
	\im[(i)] The scientist poses a question: Are all ravens black? There are two potential answers: ``Yes'' and ``No''. Those constitute the {\em competing hypotheses} of the problem. 
	
	\im[(ii)] The scientist plans to collect ravens and record their colors. A body of evidence consists of the observed colors of the sampled ravens. 
		
	\im[(iii)] The scientist assumes that counterexamples to the general hypothesis, if any, would eventually be observed were the inquiry to unfold indefinitely. This is the {\em background assumption} of the problem. 
	\ed 
	\eed  
\noindent This presentation highlights three key elements of an empirical problem in general: (i) the competing hypotheses under consideration, (ii) the possible bodies of evidence under consideration, and (iii) the background assumptions.

To clarify, the scientist could have pursued a different empirical problem.
	For instance, the background assumption could have been weakened or removed entirely; the question could have concerned statistical hypotheses; the number of competing hypotheses could have been infinite. These modifications would not affect the philosophical points I want to make but would only complicate the mathematics in use. So, I will keep the discussion simple by focusing on the raven problem as specified above, introducing more complex empirical problems only when necessary.

\begin{figure}[ht]
	\centering \includegraphics[width=0.6\textwidth]{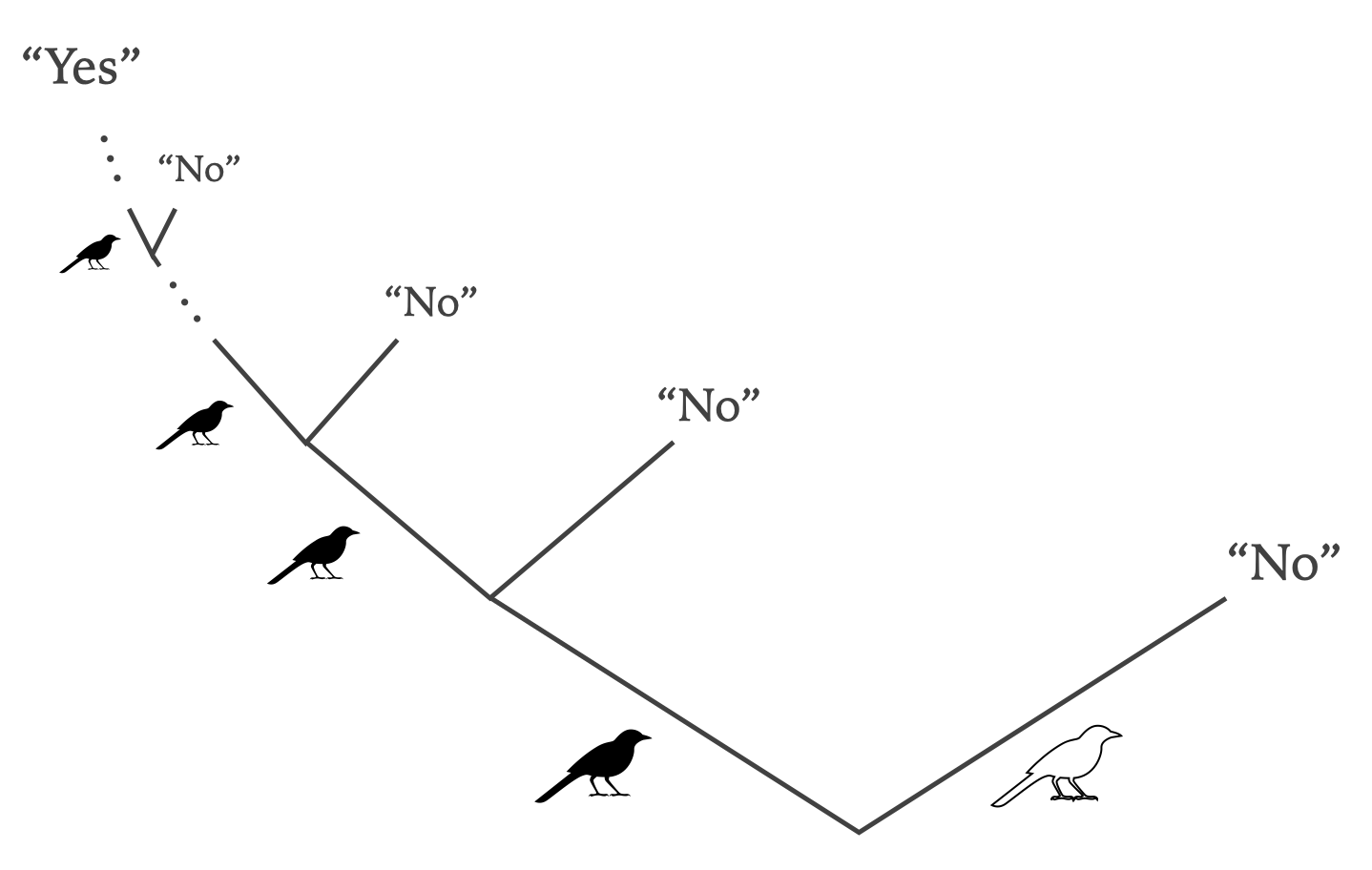}
	\caption{The raven problem, represented by a tree}
	\label{fig-tree}
	\end{figure}

The raven problem is simple enough to allow a nice visual representation, as illustrated by the tree in Figure \ref{fig-tree}.
The tree has many branches: moving upward to the left represents observing a black raven, while moving to the right represents observing a nonblack raven---a counterexample to the general hypothesis. Each node represents a possible body of evidence, while each branch represents a possible world (more or less coarse-grained). The tip of each branch is labeled with the hypothesis that is true in that world: either ``Yes, all ravens are black'' or ``No, not all are.'' Strictly speaking, some possible worlds are not represented in the tree, but that is intentional---those are precisely the worlds ruled out by the background assumption stated earlier. 

One more clarification: Although the leftmost branch is depicted as an infinite sequence of black ravens, it does not represent a world in which the scientist is immortal and observes an infinite number of ravens---that would be absurd. Instead, it simply represents a world in which all ravens are black and, hence, every raven observed {\em would} be black even if the number of observations were extended to $n$, for any finite number $n$.

So this is the setup. What normative constraints should govern the scientist's credences?

\subsection{Case Study I: Open-Mindedness vs. Regularity}\label{sec-regular}

Let me start by presenting the simplest example of a forward-looking norm that conflicts with a traditional Bayesian view---specifically, subjective Bayesianism (de Finetti 1974, Savage 1972, Lewis 1980, Skyrms 1980, van Fraassen 1989).

Shimony (1970) proposes a norm on posteriors, Open-Mindedness, which can be informally stated as follows: in any empirical problem, each of the competing hypotheses ought to have an opportunity to be most favored. This norm backward induces a constraint on (conditional) priors in the raven problem, thanks to Conditionalization: 
\oop 
	{\bf Open-Mindedness} (Local Version). In the raven problem, one's prior $\PP$ ought to be such that, for each of the two competing hypotheses $H$, there exists possible evidence $E$ such that 
	$$\PP(H \given E) ~>~ \frac{1}{2} \,.$$
\eed 


Here is the point I want to make with this example: Open-Mindedness actually imposes a distinctive constraint on priors that cannot be derived from the subjectivist view. Subjective Bayesians maintain that the correct norms on priors are exhausted by coherence norms. A key feature of coherence norms is that violating them makes one vulnerable to a certain dominance argument, such as a Dutch Book argument leading to a sure loss (de Finetti 1974) or an accuracy dominance argument (de Finetti 1974, Joyce 1998). 
Here is a notable example of such a norm:\footnote{The earliest proponent of Regularity I am aware of is Carnap (1945), who incorporates it into his objective Bayesian view. The emphasis of this subsection is on the contrast between forward-looking and subjective views, though.}
	\oop 
	{\bf Regularity}. Any proposition that is logically consistent with the available evidence, if assigned a credence at all, ought to be assigned a credence greater than $0$---the credence assigned to a logical contradiction.
	\eed 
Among the candidates for coherence norms, Regularity is perhaps the closest in spirit to Open-Mindedness. Indeed, a common motivation for Regularity is the concern that assigning a hypothesis a prior credence of $0$ would imply that it can never receive a positive posterior, no matter the evidence (Lewis 1980, Skyrms 1980). This is a forward-looking motivation for Regularity! 

Unfortunately, Regularity fails to live up to the forward-looking motivation---it is either too strong to be plausible or too weak to ensure Open-Mindedness. To see why, consider the following two cases: 

{\em Case 1}: Suppose that credences must be real numbers. Then Regularity is too strong to be plausible for a familiar reason: it rules out one of the most often used prior distributions in statistical applications---the {\em uniform} distribution on the unit interval $[0,1]$ (which can represent, for example, the set of all possible biases of a coin). The uniform distribution assigns to every closed interval $[a, b] \subseteq [0,1]$ a probability equal to its length, $b - a$. Consequently, a single point ${a}$, viewed as the degenerate interval $[a, a]$, is assigned probability zero ($0 = a - a$), violating Regularity.

\begin{figure}[ht]
	\centering \includegraphics[width=0.7\textwidth]{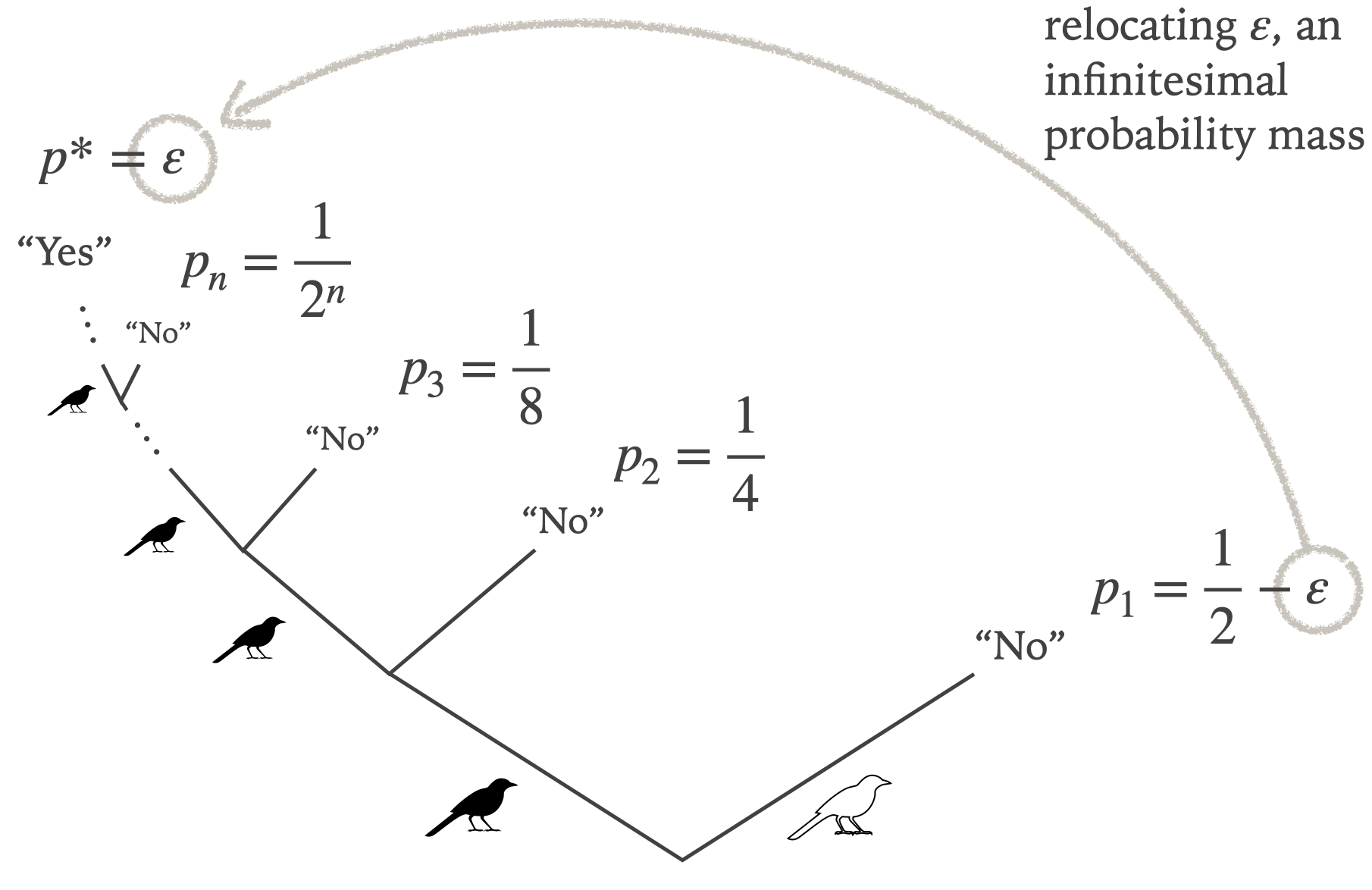}
	\caption{A regular prior with infinitesimals}
	\label{fig-regular}
	\end{figure}

{\em Case 2}: Suppose that credences are allowed to take infinitesimal values---numbers greater than $0$ but smaller than any real number. Then the hypothesis ``Yes'', which is true only at the leftmost branch of the tree, could be assigned an infinitesimal prior credence $\varepsilon$, as shown in Figure \ref{fig-regular}. If $\varepsilon$ is too small compared with the prior credences of the other branches, then its posterior credence could never reach the level of real numbers, let alone exceeding $1/2$, thereby violating Open-Mindedness. For a numerical example, see the credence distribution in Figure \ref{fig-regular} with the calculation details provided in the following footnote.\footnote
	{With the prior credences in Figure \ref{fig-regular}, the conditional credence of ``Yes'' can be calculated as follows:
	\begin{eqnarray*}
	\PP( \textrm{Yes} \given n \textrm{ black ravens in a row})
	&=& \frac{p^*}{ p_{n+1} + p_{n+2} + p_{n+3} + \cdots}
	\\
	&=& \frac{\varepsilon}{ 1/2^{n+1} + 1/2^{n+2} +  1/2^{n+3} + \cdots }
	\\
	&<& \frac{\varepsilon}{ 1/2^{n+1} } 
	\;\;=\;\; \varepsilon\,2^{n+1}
	\;\;<~\;\; 1/2 \,.
	\end{eqnarray*}
The last inequality (``$< 1/2$'') needs a small proof: Suppose for {\em reductio} that $\varepsilon\,2^{n+1} \ge 1/2$. Then $\varepsilon \ge 1/2^{n+2}$, which implies that $\varepsilon$ is not an infinitesimal---contradiction.
	}
Infinitesimal probabilities could play other interesting roles in Bayesian learning (Huttegger 2022), but this example is sufficient for the present purposes.

Thus, as shown by the discussion of Cases 1 and 2, Regularity is either too strong to be plausible or too weak to ensure Open-Mindedness. This suggests that Open-Mindedness generates a constraint on priors that goes beyond Regularity and probably all other coherence norms---the only norms that subjective Bayesians accept. If so, this provides a simple example of a forward-looking view that contradicts the subjectivist view. Forward-looking Bayesianism is a genuine alternative.

\subsection{Case Study II: Convergence vs. Countable Additivity}\label{sec-countable}


Even when a forward-looking Bayesian and a traditional Bayesian agree on certain norms, they may still disagree on which norms are more fundamental. This is precisely the idea behind the third core thesis of forward-looking Bayesianism: Fundamentality. Let me provide an example.

Some forward-looking Bayesians go beyond the virtue of being open-minded and advocate a distinctive yet controversial virtue: convergence to the truth. The core idea, roughly put, is that belief change in response to evidence should be reliable in leading to the truth across a range of possible worlds---at least when the amount of evidence becomes large enough. To illustrate, let's return to the raven problem and consider the following convergence norm:
	\oop   
	{\bf Simple Convergence}. In the raven problem, an agent's credences ought to satisfy the following condition: in every possible world $w$ compatible with the background assumption (i.e., in every branch of the tree), the posterior credence in the true hypothesis would converge to $1$ if the amount of evidence were to extend to any arbitrarily large---though still finite---quanity.
	\eed   
Applying this norm to the leftmost (all-black-ravens) world, and assuming Conditionalization, it follows that one's (conditional) prior credences ought to satisfy the following constraint: 
	\begin{eqnarray*}
	\lim_{n \rightarrow \infty}
	\PP( \textrm{Yes} \given n \textrm{ black ravens in a row}) 
	&=& 1 \,.
	\end{eqnarray*}

Some traditional Bayesians, such as Blackwell \& Dubins (1962) and Gaifman \& Snir (1982), also love convergence. They might be willing to accept Simple Convergence, but only as a norm derived from Conditionalization and certain norms on priors---such as Regularity (suitably formulated) and, especially, Countable Additivity:
	\oop 
	{\bf Countable Additivity (A Universal Version)}. It ought to be that the credence in any countable disjunction of propositions be the sum of the credences in those disjuncts, provided that those disjuncts are mutually incompatible.
	\eed 
\noindent  Under this traditional picture, although convergence is beloved as a requirement on posteriors, it is little more than icing on the cake---a desirable feature rather than a fundamental norm. 

By contrast, convergentist Bayesians---as true lovers of convergence---regard some convergence norms as more fundamental than some norms on priors, using the former to derive the latter. For example, a convergentist might take Simple Convergence as more fundamental and use it to derive a local version of Countable Additivity, that the credences assigned to the (countably many) branches of the tree in the raven problem ought to sum to $1$; a derivation is provided in the following footnote.\footnote
	 {Proof. Let $p^*$ denote the prior credence in the all-black-ravens world, i.e. the leftmost branch; and $p_{n}$ the credence in the $n$-th branch from the right. The local version of Countable Additivity can then be derived from Simple Convergence as follows:
	 \begin{eqnarray*}
&& 
	\lim_{n \rightarrow \infty} \PP( \textrm{Yes} \given n \textrm{ black ravens in a row}) \;=\; 1	
\\
&\Rightarrow &
	\lim_{n \rightarrow \infty} \left[ 
	\frac{\PP(\textrm{Yes} \;\wedge\; n \textrm{ black ravens in a row})}{\PP(n \textrm{ black ravens in a row})} 
	\right] \;=\; 1
\\[0.3em]
&\Rightarrow &
	\lim_{n \rightarrow \infty} \left[
	\frac{p^*}{1 - (p_1 + \cdots + p_n)} 
	\right] \;=\; 1
\\[0.3em]
&\Rightarrow &
	\lim_{n \rightarrow \infty} \big[ 1 - (p_1 + \cdots + p_n)\big] \;=\; p^*
\\
&\Rightarrow &
	p^* + \sum_{n=1}^{\infty} p_n \;=\; 1 \,.
\end{eqnarray*}
	 }
The point just made is not restricted to the raven problem. Some convergentist Bayesians take a step further, employing some convergence norms as reasons for accepting more global versions of Countable Additivity (Huttegger 2017, p.\,170). 

I urge traditional lovers of convergence to consider adopting the forward-looking perspective instead. Their approach, which seeks to derive convergence norms, relies essentially on Countable Additivity, as Kelly (1996, ch. 13) has observed. But Countable Additivity remains controversial as a universal requirement of coherence, with putative counterexamples such as de Finetti's lottery---a {\em fair} lottery with a {\em countable infinity} of tickets (de Finetti 1974).\footnote{But see Williamson (1999) for a reply to de Finetti.} Regardless of whether Countable Additivity can ultimately be justified as a universal norm on priors, that issue should be seen as secondary for true lovers of convergence. If a convergence property of posteriors is indeed so desirable, we may as well incorporate it into a forward-looking norm and treat it as fundamental rather than derived.

\subsection{Case Study III: Convergence and Ockham's Razor}\label{sec-nonparametric}

Let me give a more interesting example, one that contrasts forward-looking Bayesians with all traditionalists---not just subjective Bayesians, but also objective Bayesians (Carnap 1945, Jaynes 1968, Williamson 2010). If I am right, this example suggests that forward-looking Bayesianism holds the best hope for justifying an important version of Ockham's razor in machine learning and artificial intelligence.  

Consider this:
	\oop
	{\em Example: A Nonparametric Regression Problem}. 
	\op 
	\im {\em Background Assumptions}: Let $X$ be an independent variable restricted to the unit interval $[0,1]$, and $Y$ be a response variable. There is an unknown function $f^*: X \to Y$ as the target to be estimated, called {\em the true curve}, only assumed to be in a very large class of curves, the class of so-called {\em square integrable} curves, which is large enough to contain all continuous curves, and even piecewise continuous curves. (More precisely, a curve $y = f(x)$ defined on $[0, 1]$ is called square integrable if $\int_0^1 |f(x)|^2 \,dx$ is finite.) 
	\im {\em Question}: Which of those curves is the true one? 
	
	\im {\em Empirical Evidence}: A body of evidence is a finite set of data points $(x, y)$ scattered on the $XY$-plane. 
	
	\im {\em Further Background Assumptions}: Each data point $(x, y)$ is assumed to be generated in two steps. First, a value $x$ of $X$ is generated randomly from the uniform distribution on the unit interval $[0, 1]$, independently of all other data points. Then, given $X = x$, a value $y$ of $Y$ is generated through this formula: 
	$
	y \,=\, f^*(x) + \varepsilon
	$, 
	where $f^*$ is the true curve, and $\varepsilon$ is the error/noise term generated randomly from a normal distribution with zero mean and unit variance, independently of all other data points. 
	\ed 
	\eed 
\noindent Note that the true curve is not assumed to be linear or quadratic; it is not even assumed to take any particular parametric form. So the subject of this problem is known as {\em nonparametric} regression.

When a Bayesian agent tackles this curve fitting problem, they must assign prior credences over a vast space of possible curves. The statisticians Diaconis and Freedman (1998) approach this challenge from a forward-looking perspective. The idea can be captured by the following informal statement of a norm on posteriors:
	\oop 
	{\bf Convergentism for Nonparametric Regression} ({\em Informal Statement}). When tackling the problem of nonparametric regression as stated above, one's inference method for posterior assignment ought to ensure that, regardless of which curve is the true one within the considered space, there is a {\em high chance} that the resulting posterior credences will become {\em highly concentrated} around the true curve---at least in an extremely favorable evidential situation, in which the size of the dataset is an extremely large (but still finite).
	\eed   
\noindent A mathematical definition of high concentration is not needed for the point I want to make, but is included in the following footnote.\footnote
	{
	A high credence highly concentrated around a curve $f$ means assigning a high credence to a small $\varepsilon$-neighborhood of $f$. This neighborhood consists of all curves $g$ such that their distance from $f$, measured by $\int_0^1 |g(x) - f(x)|^2 \,dx$, is less than $\varepsilon$.
	}
To make this statement fully formal, we also need to quantify over all thresholds for high chance below $1$, and all thresholds for high concentration short of perfect concentration. Nonetheless, the core ideas are already present in the informal version.

Diaconis and Freedman (1998) prove a very interesting result: under Conditionalization, the above convergence norm backward induces a strong constraint on priors, embodying a version of Ockham's razor. Let me informally state their result below. 

Let $k$ be a natural number. Iteratively bisect the unit interval $[0,1]$ on the $X$-axis $k$ times, yielding $2^k$ equal subintervals. Let $M_k$, the $k$-th model, denote the class of step functions that remain constant over each of those $2^k$ subintervals and make exactly $2^k -1$ discontinuous jumps (at the junction points of those subintervals). The step function models $M_0, M_1, \ldots, M_k, \ldots$ are thus mutually exclusive; they do not jointly exhaust the entire curve space under consideration, which also contains quadratics, cubics, and any other square-integrable curves. Yet any curve in the considered space can be approximated by such step functions to arbitrary precision. Each step function model $M_k$ can be understood to have a free parameter whose value lives in a $2^k$-dimensional space. As $k$ increases, $M_k$ becomes more complex. By making $k$ large enough, you can pick a curve from $M_k$ that fits a given dataset very well---as much as you please. Now, a prior $\mathsf{Pr}$ on the considered space is said to have {\em model-wise full support} if, for any model $M_k$ construed as a subspace, $\mathsf{Pr}$ assigns a positive probability to every open region within $M_k$---or speaking informally, $\mathsf{Pr}$ spreads throughout $M_k$. Then we have:
	\opp 
	{\bf Theorem 2.3 of Diaconis \& Freedman (1998)} ({\em Informal Statement}). Let $\mathsf{Pr}$ be a probabilistic prior for the present regression problem, to be driven by data through Conditionalization, with posterior credences obeying Convergentism for Nonparametric Regression. It follows that, if $\mathsf{Pr}$ satisfies the following two properties:
		\op 
		\im[(i)] model-wise full support, 
		\im[(ii)] countable additivity,
		\ed  
	then the prior credences in progressively complex models,
$$\mathsf{Pr}(M_0), \mathsf{Pr}(M_1), \ldots, \mathsf{Pr}(M_k), \ldots \,,$$
must follow a form of Ockham's razor: dropping faster than the exponential decay $\left(\frac{1}{2^k}\right)$, and even faster than the factorial decay $\left(\frac{1}{k!}\right)$, up to a normalizing factor. 
\edd 
Let me briefly comment on properties (i) and (ii). Property (i), model-wise full support, is a very natural requirement in the context of regression, even though it may not be a universal norm. Property (ii), countable additivity, likewise may not be a universal norm; however, for practical purposes, adopting it is often prudent, since without it, statisticians have no clear way to compute posterior credences approximately using standard MCMC (Markov Chain Monte Carlo) algorithms.\footnote 
	{	
	MCMC algorithms refer to probability density functions or, more generally, Radon-Nikodym derivatives, whose existence presupposes countable additivity. Moreover, a key to justifying an MCMC algorithm is the requirement that ``the transition probability distributions must be constructed so that the Markov chain [in use] converges to a unique stationary distribution that is the posterior distribution [we want to approximate]'' (Gelman et al. 2014, p. 275). The proof of such a convergence result---like many other convergence theorems in probability theory---relies on countable additivity; see Gelman et al. (2014, ch. 12), Robert \& Casella (2004, sec. 6.6), and Meyn \& Tweedie (1993, sec. 13.3), listed here in the order of increasing technical depth.
	}
Indeed, it is standard practice in Bayesian nonparametric regression to adopt priors satisfying both (i) and (ii). In this case, as the theorem above states, Conditionalization and Convergentism for Nonparametric Regression together enforce a particularly strong form of Ockham's razor.

While the above result is only stated for regression, Diaconis \& Freedman (1998) also prove a similar result for a different inferential task: classification (Theorem 1.1). Classification can be seen as a variant of regression: the $X$-axis is extended to represent a richer space, such as a space of possible images, while the $Y$-axis is restricted to a finite set of labels, such as $1$ (for ``it's a cat'') and $0$ (for ``it's not a cat'').
	In this setting, each ``curve'' on the ``$XY$-plane'' represents a classifier---a function that assigns labels to inputs. Classification problems, along with regression problems, are central to machine learning and artificial intelligence, and they are almost always treated without any parametric assumptions in those areas.

How do traditional Bayesians fare in this case? 

First, subjective Bayesians accept only coherence norms, which impose almost no constraint on the prior credences assigned to progressively complex models, $\mathsf{Pr}(M_0), \mathsf{Pr}(M_1), \ldots$. In particular, coherence allows this infinite sequence to decay as slowly as one wishes, thus {\em permitting} defiance of Ockham's razor. 

The situation is worse for objective Bayesians. They generally favor flat priors, so they would demand that the infinite sequence of prior credences $\mathsf{Pr}(M_0), \mathsf{Pr}(M_1), \ldots$ be as flat as possible---the flatter, the better. That is, the slower the decay, the better. But this means not just permitting but {\em requiring} defiance of Ockham's razor. To dramatize: consider a very complex model, say, $M_k$ with $k =10^{100}$, and imagine defying Ockham's razor in this way: let a low prior credence be spread quite evenly from the simplest model $M_0$ up to $M_{10^{100}}$, and let a high prior credence be spread quite evenly over the extremely complex models $M_{10^{100}+1}, M_{10^{100}+2}, \ldots, M_{k}, \ldots$ for all $k \ge 10^{100} + 1$. Such a radical defiance of Ockham's razor is required by objective Bayesianism.

Upshot: traditional Bayesian approaches---both subjectivist and objectivist---seem to offer no hope of justifying Ockham's razor in nonparametric problems. Only convergentist Bayesianism offers a viable path forward.

This leaves an important message to Bayesians: the foundational work in machine learning and artificial intelligence is largely driven by the need for making only weak background assumptions---particularly, by removing parametric assumptions. These foundational developments have been predominantly situated within a frequentist, non-Bayesian framework (Devroye et al. 1996, Gy\"{o}rfi et al. 2006, Shalev-Shwartz et al. 2014, Mohri et al. 2018).
		If Bayesians wish to catch up in this crucial area, the key lies in nonparametric Bayesian statistics (Wasserman 1998, Ghosal et al. 2017), whose standard practice follows convergentist norms. 

For these reasons, I believe the convergentist approach represents the best---if not the only---hope for Bayesians seeking a solid foundation of machine learning and artificial intelligence. This explains why it is crucial to defend the plausibility of forward-looking Bayesianism in general, and convergence norms in particular. This is the central task of the remainder of the paper.





\section{A General Worry from Default Bayes}\label{sec-default-bayes}

Forward-looking norms might seem quite strange from the traditional, default Bayesian perspective, which can be formulated as follows:
	\oop  
	{\bf Default Bayesian View, Part 1: Initialization}. One's inquiry always begins with some initial opinions.
	
	{\bf Default Bayesian View, Part 2: (Na\"{i}ve) Probabilism}. Those initial opinions are represented as degrees of belief, required to follow at least the norm of Probabilism, forming a probability measure. This probability measure assigns prior credences to all propositions relevant to the inquiry, including competing hypotheses $H$, possible bodies of evidence $E$, and their Boolean combinations, such as conjunctions $H \wedge E$.
	\eed  
	\oop  
	{\bf Default Bayesian View, Part 3: (One-Sided) Conditionalization}. There is no need to impose a norm directly on posterior credences. For the prior credences from Part 2 already determine what the posterior credences should be by the Conditionalization norm, which requires the following, where $E$ is the available evidence:
	$$
	\text{Posterior credence in } H 
	~=~ 
	\frac{\text{Prior credence in } (H \wedge E)}{\text{Prior credence in } E} \,.
	$$ 
	\eed  
\noindent Note that the two prior credences on the right-hand side of the equation are already provided in Part 2---they are assigned by the initial probability measure.

The default perspective suggests a picture of ``living (entirely) in the moment'' without ``thinking ahead.'' The idea is to concern ourselves with posterior credences only when new evidence is actually acquired. Until then, it seems sufficient to regulate prior credences solely through synchronic norms, focusing only on the present. Once prior credences are properly set, they will be updated through Conditionalization, which automatically ensures the proper formation of posterior credences as new evidence arrives. If so, there appears to be no role for forward-looking norms to play. This default perspective appears to be implicitly assumed by many Bayesians, which may help explain why forward-looking norms are rarely discussed---whether in defense or in critique.

I reply that the default Bayesian view is incorrect. It corresponds to a convenient textbook presentation, serving as a crude approximation to the Bayesian views that have been seriously defended by Bayesians. Specifically, this view assumes a version of Probabilism that is too strong to be plausible; that is, Part 2 of the default view is problematic, and it is presupposed by Part 3. Once a weaker, better version of Probabilism is in place, an epistemic role for forward-looking norms naturally emerges---or so I will argue below.

\subsection{Which Version of Probabilism?}\label{sec-extendable}

There is a grain of truth in the default Bayesian view, namely Part 1: every inquiry begins with some initial opinions---the opinions that are at least provisionally taken for granted at the start of one's inquiry. However, one is not required to assign a prior credence to every considered hypothesis $H$, every possible body of evidence $E$, and every conjunction $H \wedge E$. Probabilism does not actually demand that much---or at least, the right version of Probabilism does not require that much.

The reason is straightforward, as already pointed out by earlier proponents of Probabilism such as de Finetti (1974, sec. 3). Probabilism is meant to be a coherence norm, specifying how one's credences ought to be structured to avoid incoherence. But coherence is a matter of fitting credences together nicely---it does not dictate which specific propositions must be assigned credences.
	Thus, if Probabilism is truly a coherence norm, it should only specify {\em which combinations of credences} are impermissible, rather than {\em which propositions} one must assign credences to. It does not require one to assign a credence to any given hypothesis $H$, evidence $E$, or their conjunction $H \wedge E$, although it does prohibit certain combinations of credences, such as assigning a higher credence to $H \wedge E$ than to $H$.
More generally, a better version of Probabilism states the following:
	\oop {\bf Probabilism} ({\em Extendability Version}). One should have a combination of credence assignments that can be {\em extended} to a probability measure, possibly by {\em forming} additional credences. 
	\eed 
It is also worth examining whether the major arguments for Probabilism, when carefully formulated, support only the extendability version rather than the default version. The answer seems to be yes---both for the Dutch Book argument (de Finetti 1974) and for the accuracy dominance argument (Pettigrew 2016).

Once we adopt the extendability version of Probabilism, forward-looking norms emerge as playing an important normative role, to which I now turn.


\subsection{Credence Formation vs. Revision}\label{sec-formation}

Imagine, from a first-person perspective, that we are agents tackling the raven problem (as specified in Section \ref{sec-raven}). We take for granted the background assumption $A$: that either all ravens are black or a counterexample would eventually be observed if the inquiry were to continue indefinitely.
	So, granted, every inquiry begins with some initial opinions (Part 1 of the default Bayesian view), and in this case, taking $A$ for granted is one such initial opinion. But now suppose that this attitude toward $A$ is the only initial opinion we hold at the start of the inquiry.
		This initial opinion---only taking $A$ for granted---can be given a particularly simple modeling: it is just the state of assigning credence of $1$ to $A$ while assigning no credence to any other propositions. Caveat: this assignment of credence $1$ is subject to subsequent revision, as Titelbaum (2013) suggests; that would amount to abandoning $A$ as the background assumption. One more caveat: I adopt this simple model of taking $A$ for granted for the sake of convenience; there is an alternative model using primitive conditional credences.\footnote
			{Following Salmon (1990) and Wenmackers \& Romeijn (2016), the attitude of taking $A$ for granted may be modeled as the state of only assigning conditional credences given $A$, where the concept of conditional credences is primitive.}

The state of having only that single credence---credence $1$ in $A$---conforms to Probabilism in its extendability version. From this starting point, we as the agents can then deliberate on what additional prior credences to form before acquiring new evidence. This process concerns the {\em formation} of new prior credences, rather than the {\em revision} of existing credences. Only the latter is governed by Conditionalization.

The formation of additional prior credences is guided at least by Probabilism (again, in its extendability version). This means that the newly formed prior credences, together with the existing credence $1$ in $A$, should still allow for the possibility of forming further prior credences to make up a probability measure. In addition to Probabilism, further norms may govern the formation of prior credences.

Here is the key point: The stage of forming prior credences is where forward-looking norms have an important role to play. 
	Take Open-Mindedness as an example. If we are not careful, we might form prior credences that would prevent a hypothesis from ever being most favored, violating Open-Mindedness (recall the example in Section \ref{sec-regular}). 
Thus, Open-Mindedness---along with other forward-looking norms---can actively shape the deliberative stage of forming prior credences. Forward-looking norms do not merely encourage {\em thinking ahead}; they are used to {\em work backward} and thus can be valuable {\em in the moment}.


So, while it may seem that both Conditionalization and forward-looking norms are meant to govern posteriors, they actually play distinct roles in an inquiry.
	\op 
	\im Conditionalization governs the revision of existing credences in response to new evidence.
	\im Forward-looking norms, by contrast, govern the formation of additional prior credences---albeit indirectly, by directly constraining posteriors and using Conditionalization as a link between priors and posteriors.
	\ed 
This is an important epistemic role that forward-looking norms can play.

\subsection{Generalization}\label{sec-generalize}

The above discussion is intended only as a crude illustration of an epistemic role of forward-looking norms. Now, let me gesture at possible refinements.

Like the raven problem, an empirical problem in general is specified by at least three key components:
\op
\im[(i)] A set $\mathscr{H}$ of competing hypotheses, representing potential answers to a given question.
\im[(ii)] A space $\mathscr{E}$ of possible bodies of evidence.
\im[(iii)] A body of background assumptions, $\mathscr{B}$, representing the opinions taken for granted in the problem.
\ed
The correct mathematical modeling of the background assumption $\mathscr{B}$ remains open to discussion. Recall the simple modeling adopted above: it treats $\mathscr{B}$ as a partial function that does nothing but assign a credence of $1$ to a single proposition $A$, which is the conjunction of all propositions taken for granted. This idea has as equivalent formulation that can be easily generalized. That is, the background assumption $\mathscr{B}$ can be equivalently represented as a candidate pool of permissible priors---specifically, the set of all probability measures that assign a credence of $1$ to $A$. Forward-looking norms, along with other possible normative constraints, serve to narrow this pool of candidates.

Now we can generalize by allowing the background assumption $\mathscr{B}$ to be any set of ``priors'', pending an account of the correct mathematical modeling of ``priors'':
	\op 
	\im Priors as probability measures (which only output sharp probability values)? 
	\im Or priors as functions that output intervals of probabilities? 
	\im Or priors as sets of probability measures? 
	\im Or priors as something even more general?
	\ed  
No matter how we fix the mathematical modeling of priors, my point remains the same: the background assumption of an empirical problem still serves as a candidate pool for permissible priors. Moreover, forward-looking norms still have an important epistemic role: they help narrow the candidate pool of permissible priors, guiding the formation of credences.


For the sake of simplicity, the following discussion will proceed under the simple modeling: priors as probability measures, and background assumptions as certainties.

\section{Another General Worry: Stuck with Your Prior}\label{sec-ought-to-be}

There is another general worry for forward-looking Bayesians. It actually embodies a sentiment that subjective Bayesians often hold against the objectivist view. Inspired by Joyce's (2011, sec. 2.3) presentation of that worry, I formulate it as follows:
	\oop  
	{\bf The ``Stuck with Your Prior'' Argument.} Consider the following scenario: An agent satisfies Probabilism by adopting a probability measure to represent her prior belief state. Now, suppose that her prior credences violate a certain norm $N$---for example, a norm requiring a flat prior.
		In this case, the agent should not immediately adjust her prior credences to conform to $N$, because doing so would violate Conditionalization. Indeed, Conditionalization prohibits credence change in the absence of new evidence.
	Hence a counterexample to $N$.
	\eed  
\noindent The idea of this argument, which may be called ``Stuck with Your Prior'', is quite general. Examples of the norm $N$ are not limited to objectivist norms such as the Principle of Indifference (which requires a flat prior), but can also be forward-looking norms. 

Here is my reply: this argument proves too much. The list of examples for $N$ can be extended further to include {\em any} norm not entailed by Probabilism and Conditionalization---such as Regularity and the Principal Principle (a norm that link personal credences to physical chances). 
Since this argument would then rule out almost all Bayesian views, it is too strong to be correct. It remains to give a diagnosis of where it goes wrong.


The ``Stuck with Your Prior'' argument starts by envisioning a scenario of normative dilemma, where an agent with a probabilistic prior faces a forced choice: they must violate either Conditionalization or some other norm $N$. But which one to violate?
	The argument implicitly assumes that such a normative dilemma must always be resolved in favor of Conditionalization---that is, Conditionalization holds normative priority over $N$. However, this presupposition is far from obvious and requires a separate justification. 

Even if there is an argument for the priority of Conditionalization, it still does not undermine the normative status of $N$, provided that $N$ is properly understood. To see this clearly, let's distinguish two types of norms, traceable to Casta\~{n}eda's (1970) work in deontic logic. Consider this: ``There ought ideally to be no war.'' This norm expresses an ideal state---a state that the world ought ideally to be in. As such, it is an {\em ought-to-be norm}, and might not provide guidance on the issue of what we ought to do in a less-than-ideal situation. Think about Churchill's situation right before the outbreak of World War II: We are given the choice between war and dishonor; if we choose dishonor, we will still have war. In this case, the norm ``there ought ideally to be no war'' offers no guidance on what we should do---this norm is to be violated anyway.  
	Now, what should we choose? This is where a different type of norm comes into play: an {\em ought-to-do} norm, designed to guide us through non-ideal situations.\footnote
		{
		The distinction between ought-to-be and ought-to-do norms is widely discussed in the broader context of normative studies, including deontic logic (Casta\~{n}eda 1970; Horty 2001: sec. 3.3 and 3.4) and metaethics (Broome 1999; Wedgwood 2006; Schroeder 2011).
		} 

Bayesian norms are often understood as norms of {\em ideal} rationality. For example:
	\op 
	\im Probabilism asserts that one's credences ought ideally to be probabilistic.
	
	\im Conditionalization asserts that one's priors, posteriors, and evidence ought ideally to be related in a certain way.

	\im Open-Mindedness (as a forward-looking norm) states that one's posteriors ought ideally to satisfy certain conditions.
	\ed 
These norms together outline an ideal epistemic state---a condition that one ought ideally to be in. As such, they function as ought-to-be norms rather than ought-to-do norms, and do not imply any guidance on what to do in less-than-ideal situations---such as the normative dilemma described in the ``Stuck with Your Prior'' argument. In this case, an agent with a probabilistic prior finds herself forced to violate either Conditionalization or some other norm $N$.
	Then, to insist that this agent ought to violate $N$ rather than Conditionalization is to maintain the priority of Conditionalization as an ought-to-do norm. However, such a stance goes beyond the two familiar Bayesian norms---Probabilism and Conditionalization as ought-to-be norms.

Some Bayesians have already resisted the priority of Conditionalization in non-ideal situations. Reflecting on the practice of Bayesian statistics, de Finetti \& Savage (1972) propose the following guideline:  
	\oop 
	{\bf An Ought-to-Do Norm for Dealing with New Theories.} In the non-ideal situation where one formulates a new hypothesis and decides to take it seriously, one ought to reassign credences in the old and new hypotheses as if \op \im[(i)] at the beginning of one's inquiry, one had formulated the new theory alongside the old ones and assigned them prior credences in conformity with Probabilism and any other correct norms on prior credences, \im[(ii)] up until now, one had changed one's credences by, and only by, Conditionalization. \ed 
	\eed  
\noindent This norm requires the agent to reassign credences even when there is no new evidence, rejecting the priority of Conditionalization as an ought-to-do norm.

Interestingly, the as-if clause (ii) suggests that de Finetti and Savage actually value Conditionalization highly as an ought-to-be norm, which points to a normative ideal.
	Thus, in a somewhat surprising twist, it is precisely the plausibility of Conditionalization as an ought-to-be norm (governing ideal situations) that undermines its priority as an ought-to-do norm in the non-ideal situation of new theories.
Perhaps this is not surprising after all. Our firm belief that there {\em ought to be} no war might lead us to conclude that we {\em ought to do} something drastic---perhaps even launching a war to end all wars. 
	A principle as an ought-to-be norm can undermine its priority as an ought-to-do norm.

To sum up, the ``Stuck with Your Prior'' argument relies on two presuppositions that Bayesians need not endorse:
\op
\im First, it presupposes that Probabilism and Conditionalization are not just ought-to-be norms but also ought-to-do norms.
\im Second, it presupposes that, as ought-to-do norms, Probabilism and Conditionalization hold the highest priority, overriding any other norm $N$ that conflicts with them in less-than-ideal situations.
\ed
It is precisely these two presuppositions that make the ``Stuck with Your Prior'' argument prove too much, creating an illusory threat not just to forward-looking Bayesians, but to almost all Bayesians, whether objective or subjective.
	Thus, at least one of these two presuppositions should be rejected---though my recommendation is to reject both.




\section{A Worry for Convergentist Bayes: It's Irrelevant!}\label{sec-death-long}


While the previous two sections addressed general worries about any forward-looking norms, my preferred version of forward-looking Bayesianism---the convergentist approach---faces its own distinct challenges. One familiar objection is this: We are all dead in the long run, so who cares about convergence? The concern is that convergence, or any epistemic criterion that appeals to the long run, seems to be epistemically irrelevant (Carnap 1945, Friedman 1979). Before responding to this objection, let me first clarify its source.

\subsection{A Decision-Theoretic Interpretation?}

A convergence norm is often dismissed as irrelevant, likely because it is (mis)understood as prescribing poor decision theory for an agent. To see why, let me restate a convergence norm discussed earlier, now annotated with highlights $(a)$ and $(b)$, which will soon be our focus:
	\oop   
	{\bf Simple Convergence}. In the raven problem, an agent's credences ought to satisfy the following condition: in every \uline{possible world $w$ compatible with the background assumption}$_{(a)}$ (i.e., in every branch of the tree), the \uline{posterior credence in the truth would converge to $1$}$_{(b)}$ if the evidence were to extend to any arbitrarily large but still finite amount.
	\eed   
\noindent This norm sounds like prescribing a decision theory for the agent. It appears to suggest that an agent tackling the raven problem should construct a decision table, where:
	\op 
	\im The rows correspond to different options as alternative plans for forming posterior credences in response to evidence.
	
	\im The columns correspond to the possible worlds $w$ as indicated by $(a)$.
	
	\im The cells specify outcomes, including especially outcomes in the remote future, concerning the agent's posterior credences in the long run, as indicated by $(b)$.
	\ed 
However, this would be a bad decision theory for the agent---who cares about those remote outcomes? If we were to fully specify these long-run outcomes, we would inevitably have to include this crucial fact: the agent's death in the long run. This, I suspect, is a major source of worry for convergence norms---they seem to anchor epistemic norms to a future outcome with limited significance.

Even worse, this decision-theoretic construal of convergence norms is reinforced by the way many convergentists describe their own views. 
	For example, Reichenbach (1938) frames his convergentist epistemology as a {\em pragmatic} solution to Hume's problem. An intellectual heir to Reichenbach's thought is formal learning theory; some proponents, such as Schulte (1999), describe it as a {\em means-ends} epistemology.
Whether or not these convergentists actually have a decision theory in mind, their presentations can easily mislead readers into thinking that convergence norms have to function decision-theoretically---as if agents must decide what beliefs to hold and what inferences to draw based on a decision table with long-run outcomes.
	That would be poor decision-making indeed. 

I urge convergentists to avoid terms like `pragmatic' and `means-ends', as these can easily mislead readers into interpreting convergence norms through a decision-theoretic lens. In what follows, I will propose a better way to understand convergence norms.

\subsection{A Thought-Experiment Interpretation}\label{sec-thought-experiments}

If I am right, convergentists are best understood as simply engaging in a standard methodology in epistemology: doing thought experiments. Here is a familiar example:
\oop  
{\em Thought Experiment: Brain in a Vat}. 
\\[.5em]
Imagine a possible scenario in which an agent is a brain in a vat. A mad scientist (or evil demon) manipulates the agent's perceptual experiences through electrodes connected to the brain, crafting these experiences to perfectly mimic those of an ordinary person in an ordinary environment.
	As a result, the agent forms the belief that she has hands---just as an ordinary person would.
		Now, the key question: Is this belief justified?
\eed   
\noindent This scenario may seem unrealistic, yet it remains highly relevant to epistemology. In fact, it seems plausible to judge that the agent's belief is justified, and this judgment---intuitive or carefully considered---has been used as a counterexample to certain theories of justified belief, such as the original version of reliabilism (Goldman 1979). This, in turn, has motivated refinements, such as the ``normality'' version of reliabilism (Leplin 2007).
	The unrealistic nature of this thought experiment has rarely, if ever, been considered a reason to dismiss its use in epistemology. Keep this point in mind.

Now, let's consider another example:
\oop  
{\em Thought Experiment: Big Data}. 
\\[.5em]
Imagine a possible scenario in which an agent is tackling the raven problem. The true hypothesis turns out to be: ``Yes, all ravens are black.'' The agent has gathered a large body of evidence, expressed as: ``I have seen $n$ ravens, and they all are black,'' where $n$ is a very large but finite number. 
		In this scenario, should the agent has a high posterior credence in the truth, ``Yes'', given the available evidence?
\eed  
This, too, is an unrealistic thought experiment. It describes an evidential situation that is unrealistically favorable---a case of big data. Nonetheless, it is natural to judge that the epistemic question posed should be answered affirmatively:
	\oop  
	{\bf A Judgment on Big Data}. The agent's credence in the truth should be high, say greater than $0.9$, in such an evidentially favorable situation with a large sample size $n$. 
	\eed  
\noindent This epistemic judgment seems quite plausible---and in fact, the larger $n$ is, the more plausible it becomes. That is, as the dataset grows larger and more unrealistic, the evidential situation becomes increasingly favorable, making this epistemic judgment even more compelling.

As far as I know, no epistemologist analyzing the Brain-in-a-Vat thought experiment infers irrelevance from unrealism. I propose that we adopt the same attitude toward the Big-Data thought experiment---it should not be dismissed as irrelevant simply because it is unrealistic.

Now, let's try to refine our epistemic judgment about the Big-Data thought experiment, and generalize it as appropriate.
	A tentative formulation might state that, in the Big Data scenario, the posterior credence in the true hypothesis ought to be high, at least when $n$ is sufficiently large. That is, the following ought to hold:
	\oop 
	{\bf Version 1: A Vague Condition}.
	\\[0.5em]
	For any $n$ that is sufficiently large,
	\\ $\PP( \textrm{Yes} \given n \textrm{ black ravens in a row}) > 0.9$.
	\eed 
\noindent This condition is quite vague---how large must $n$ be to count as sufficiently large? 
	Fortunately, we can formulate a weaker but more precise condition. Suppose, plausibly, that there exists some threshold number $N$ that qualifies as sufficiently large in the scenario, and that any larger number also qualifies.
		Then, the vague condition above implies the following more precise counterpart, which ought to hold:
	\oop 
	{\bf Version 2: Weakened but Sharpened}.
	\\[0.5em]
	There exists a finite number $N$ such that, for any $n \ge N$, 
	\\ $\PP( \textrm{Yes} \given n \textrm{ black ravens in a row}) > 0.9$.
	\eed 
\noindent Since this judgment is fairly weak, it should still hold even if the threshold for high credence is raised from 0.9 to any number less than 1. This motivates a stronger judgment, asserting that the following ought to be the case:
	\oop 
	{\bf Version 3: Strengthened and Still Sharp}.
	\\[0.5em]
	For any threshold of high credence $1-\varepsilon$ less than $1$,
	\\there exists a finite number $N$ such that, for any $n \ge N$, 
	\\ $\PP( \textrm{Yes} \given n \textrm{ black ravens in a row}) > 1-\varepsilon$.
	\eed 
\noindent But the above statement is simply the formal definition of Bayesian convergence to the truth, which can be concisely expressed as follows:
	\oop
	{\bf Version 4: Restated in a Mathematical Notation}.
	\\[0.5em]
	$\lim_{n \rightarrow \infty}
	\PP( \textrm{Yes} \given n \textrm{ black ravens in a row}) \;=\; 1$. 
	\eed
\noindent We have just witnessed an application of a common methodology in epistemology, exemplified by the Big-Data thought experiment and the progression of epistemic judgments about it---from the vague formulation (Version 1) all the way to the standard expression of convergence (Version 4). At no point does this process involve decision theory, let alone bad decision theory. From here, only a slight generalization is needed to arrive at the norm of Simple Convergence.

At this point, skeptics of convergentism might still have a lingering worry: Even if the Big-Data scenario---with an extremely large $n$---can be epistemically relevant in the same way that the Brain-in-a-Vat scenario is, there is still something lacking. It still seems irrelevant to concrete, realistic cases. The concern is that Big Data only pertains to unrealistic cases and thus has {\em no implications} for an agent dealing with a realistic dataset. Or so the worry goes.

A reply is available using some recourses mentioned above: the charge that there are no implications for realistic cases is misleading.
	Conditionalization imposes a {\em cross-scenario} normative constraint---it links scenarios with a small realistic sample size and those with a large unrealistic sample size. Because of this joint constraint, a norm applying in one scenario may induce a norm in the other. Just recall the seesaw analogy presented in the introductory section. 
		Indeed, we have seen that convergence norms can lead to a substantive constraint on priors---such as a local version of Countable Additivity (Section \ref{sec-countable}, and a strong form of Ockham's razor (Section \ref{sec-nonparametric}).	
This means that convergence norms do have implications for the short run---even for the present, shaping one's prior credences.

To be sure, the idea of convergence can be misused. So, it is important to be aware of bad applications in order to avoid them.
	Consider the probability distribution of the relative frequency of heads in $n$ coin tosses. According to the central limit theorem, this distribution converges to a normal distribution as $n$ increases indefinitely. However, if all we have is this asymptotic result, we are not justified in inferring that the distribution is already close to normal given the sample size we actually have.
		This would be an error that conflates a long-run property with a short-run one---a conflation of long-run convergence with actual approximation.
Yet convergentist epistemology does not conflate long-run convergence with actual approximation.

Instead, convergentist epistemology is best understood as starting with thought experiments about possible scenarios in which we have big data. Norms about those unrealistically favorable scenarios are then elicited and refined, becoming convergence norms. Those convergence norms are in turn conjoined with Conditionalization as a cross-scenario principle to derive some norms applicable to our actual situations, such as norms on prior credences. Such a cross-scenario principle is a crucial bridge, connecting the long run to the short run---the actual scenario. This, I submit, is an unproblematic way to use the idea of convergence in epistemology, making it highly relevant to our epistemic concerns.


\section{More Worry for Convergentist Bayes: Too Messy!}\label{sec-achievabilist}

Another reason why some Bayesians might resist the convergentist approach is that it may appear messy and patchy. Different convergentists adopt different definitions of convergence to the truth, leading to norms that seem to serve only as local norms, applicable to restricted contexts.
	For instance, chances are not mentioned at all in the mode of convergence used in the context of the raven problem in Section \ref{sec-countable}, whereas a stochastic mode of convergence is employed in the context of the curve-fitting problem in Section \ref{sec-nonparametric}.
So, the convergentist approach might seem messy, which stands in stark contrast to one of the key reasons for which many people embrace the Bayesian framework: its systematicity, elegance, and arguably powerful explanatory scope across a wide range of epistemic scenarios.
	Thus, the worry is that the convergentist approach risks disrupting this beautifully structured Bayesian framework.

Here is my reply: the convergentist approach has recently been systematized at least in the epistemology of beliefs (Lin, 2025), and it is easy to carry it over to the epistemology of credences. The idea is straightforward:
	\oop 
	{\bf The General Idea of Achievabilist Convergentism}. Different definitions of convergence to the truth reflect higher or lower epistemic standards, and the choice of which standard to adopt in a given problem context is guided by a unifying principle: in each problem context, we should strive for the highest achievable standard.
	\eed 
So, the mode of convergence in use---as an epistemic standard---changes from one problem context to another only because we ought to systematically set the highest achievable standard in each problem context.  

Let me illustrate with a simple hierarchy of modes of convergence, ordered from high to low---with intuitive explanations to follow immediately, and definitions relegated to footnotes:
\oop 
$$\begin{array}{l}
	\quad\quad\quad\quad\quad  \textbf{An Example of a Hierarchy}
\\[1em]
	\text{\uline{Convergence} to the Truth with Identification}
\\[0.5em]
	\;\;\downarrow \textsf{lower the bar}
\\[0.5em]
	\text{\uline{Stochastic Convergence} to the Truth with \uline{Identification}}
\\[0.5em]
	\quad\quad\quad\quad\quad\quad\quad\quad\quad\quad\quad\quad\quad\quad\quad\quad\quad\quad\quad\,\downarrow \textsf{lower the bar}
\\[0.5em]
	\text{Stochastic Convergence to the Truth with \uline{Approximation}}
\end{array}$$ 
\eed 
\noindent The highest mode of convergence imposes a demanding requirement: in every possible world compatible with the background assumptions of the empirical problem at hand, one's posterior credence in the truth (among the competing hypotheses) must rise and remain high as the amount of evidence increases indefinitely.\footnote 
	{
	Here is a more precise definition: an inference method for assigning posterior credences in an empirical problem is said to achieve convergence to the truth with {\em identification} if in every possible world compatible with the background assumptions,  the posterior credence in the true hypothesis converges to $1$ as the amount of evidence increases indefinitely.
	}
This is the mode of convergence employed in the raven problem (Section \ref{sec-countable}). 

The second mode lowers the bar: it is not the credence in the truth, but the {\em chance} of a high posterior in the truth, that is required to rise and stay high, allowing for the unlucky but objectively unlikely event that the posterior in the truth always stays low.\footnote
	{
	Here is a more precise definition: an inference method for assigning posterior credences in an empirical problem is said to achieve {\em stochastic} convergence to the truth with {\em identification} if, in every possible world compatible with the background assumptions, for any threshold of high probability $1-\varepsilon $ less than $1$, the chance that $\big[\,$the posterior credence in the true hypothesis is greater than $1-\varepsilon$$\,\big]$ converges to $1$ as the amount of evidence increases indefinitely. This mode of convergence is also known as {\em Bayesian consistency for hypothesis testing}.
	}

The third mode lowers the bar even further: the desideratum is not a high chance of a high credence in exactly the truth, but only a high chance of a high credence {\em highly concentrated} around the truth. That is, it concerns a high chance of a high credence assigned to a small open interval covering the truth---or more generally, a small open neighborhood of the truth in a possibly high-dimensional space.\footnote 
	{
	Here is a more precise definition: an inference method for assigning posterior credences in an empirical problem is said to achieve {\em stochastic} convergence to the truth with {\em approximation} if, in every possible world compatible with the background assumptions, for every threshold of high probability $1-\varepsilon $ less than $1$, for every (true) disjunction $D$ of hypotheses being an open neighborhood of the true hypothesis, the chance that $\big[\,$the posterior credence in the (true) disjunction $D$ is greater than $1-\varepsilon$$\,\big]$ converges to $1$ as the amount of evidence increases indefinitely.
		This mode of convergence is known as {\em Bayesian consistency for estimation} in statistics. 
	}
This third mode of convergence is the one employed in the problem of nonparametric regression (Section \ref{sec-nonparametric}), in which the two higher modes of convergence are unachievable. In contrast, the highest mode of the three is achievable in the raven problem. 

The above is intended only as a very crude approximation to the correct hierarchy of modes of convergence. We can, for instance, incorporate rates of convergence, or even require a uniformly high convergence rate across all possible worlds under consideration. There are many more possible modes of convergence to explore. 
		Regardless of which specific hierarchy is the correct one, convergentist Bayesians can embrace this thesis:
	\oop 
	{\bf The Achievabilist Thesis for Convergentist Bayesianism}. If an agent tackles an empirical problem $\mathscr{P}$ and adopts an inference method for assigning posterior credences in response to evidence, then the inference method ought to achieve the highest mode of convergence to the truth achievable in problem $\mathscr{P}$---pending a specification of the correct hierarchy of modes of convergence. 
	\eed
Caveat: this statement is only a first approximation. Several complications may arise:
	\ope 

	\im {\em A Nonlinear Hierarchy}: The correct hierarchy of modes of convergence may not be linear but instead a partial order, where two incommensurable standards exist---neither higher than the other. If so, there might not be a uniquely highest achievable standard in a given empirical problem.

	\im {\em No Single Correct Hierarchy}: There may be no single ``correct'' hierarchy, but instead a context-dependent or pluralistic structure.

	\im {\em A Generalized Modeling of Priors}: The above definitions of modes of convergence may need to be generalized if we adopt a broader modeling of Bayesian priors and background assumptions, as briefly discussed in Section \ref{sec-generalize}.
	\ede 
In any of these cases, the statement of the thesis would need to be modified accordingly.

I hope that this is enough to demonstrate that there need not be anything messy or patchy in the convergentist approach---it is indeed possible to unify the many convergence norms that have been postulated for separate problem contexts.




\section{Closing}

Forward-looking Bayesianism offers a distinctive approach to the problem of the priors: instead of only directly imposing norms on priors, it also addresses the problem of the posteriors and then works backward to derive normative constraints on priors. The goal of this paper is to spark interest in this approach and encourage further exploration of its foundations, implications, and potential advantages.

I began by explaining how the forward-looking approach differs significantly from the more traditional Bayesian views. 
	First, it provides norms stronger than some requirements of coherence found in subjective Bayesianism (Section \ref{sec-regular}).
		Second, it takes a very different stance on which norms are fundamental versus derived, even disagreeing with the traditional Bayesians who are already sympathetic to convergence (Section \ref{sec-countable}).
			Perhaps most strikingly, in contrast to both subjective and objective Bayesian views, the forward-looking approach (in the convergentist form) appears to offer the only viable path to justifying Ockham's razor in nonparametric problems (Section \ref{sec-nonparametric}).

I have also defended forward-looking Bayesianism as a general framework. 
	It is perfectly compatible with Probabilism and Conditionalization, provided that these two familiar Bayesian norms are properly construed---that is, understood as {\em extendability} and {\em ought-to-be} norms (Section \ref{sec-ought-to-be}).
		Moreover, although forward-looking norms directly govern posterior credences, they also have an important epistemic role in the present, not just in the future. Specifically, they can guide the {\em formation} of prior credences, even though the {\em revision} of prior credences remains to be governed by Conditionalization (Section \ref{sec-formation}).

While the framework of forward-looking Bayesianism allows for different implementations, my preferred choice is convergentist. The main reason is that it seems to provide the only viable Bayesian justification of Ockham's razor for nonparametric problems---the primary concerns in the very foundations of machine learning and artificial intelligence (Section \ref{sec-nonparametric}).
	I defended convergentist Bayesianism against the usual Keynesian objection ``we are all dead in the long run.'' The key is that convergentist epistemology is best understood not as a form of decision theory, but rather as following the standard methodology of thought experiments in epistemology (Section \ref{sec-thought-experiments}).
		There was also the concern that evaluative standards in convergentism are messy and patchy. In response, I proposed a way to systematize it by going {\em achievabilist}: the epistemic standard in use changes from context to context only because we ought to systematically set the highest achievable one in each context (Section \ref{sec-achievabilist}).

Much more remains to be done. Let me highlight some notable and particularly urgent tasks:
	\op 
	\im {\em Further Case Studies}: Additional case studies are needed to assess whether forward-looking Bayesianism, and convergentist Bayesianism in particular, contribute to a better Bayesian philosophy of science---and a viable Bayesian foundation of machine learning and artificial intelligence.
	
	\im {\em Bridging Frequentism and Bayesianism in Statistics}: We have seen an interesting mode of convergence: {\em stochastic (chancy)} convergence of {\em credences} to the truth. It integrates ideas from two seemingly opposing schools in the foundations of statistics: frequentism and Bayesianism. This suggests that the two schools should not be treated as a strict dichotomy, as traditionally construed (Hacking 1965/2016), but rather as compatible---and potentially capable of forming a productive alliance. This alliance needs further exploration and critical examination (see Lin (2024) for an initial attempt).

	\im {\em Alliance with Accuracy-First Bayesianism}: The pursuit of accuracy---i.e., high credence in the truth---is an epistemic goal shared by many. It is valued not only by convergentist Bayesians but also by those who defend familiar Bayesian norms using accuracy-first arguments (Joyce 1998; Pettigrew 2016; Konek \& Levinstein 2019). Thus, an alliance between these two groups may be possible and is worth investigating.	
	\ed 
I look forward to exploring these developments, though they will have to be reserved for future work.


\section*{Acknowledgements}

I thank Frederick Eberhardt, Christopher Hitchcock, Boris Babic, and Eddy Keming Chen for discussions. I am especially indebted to Alan H\'{a}jek for his continued encouragement and many helpful conversations about this project.

\section*{References}

\begin{description}
\im Belot, G. (2013). Bayesian orgulity. \textit{Philosophy of Science}, 80(4), 483-503.

\im Blackwell, D., \& Dubins, L. (1962). Merging of opinions with increasing information. \textit{Annals of Mathematical Statistics}, 33(3), 882-886.

\im Broome, J., 1999, ``Normative Requirements'', {\em Ratio}, 12(4): 398-419.

\im Carnap, R. (1945). On inductive logic. \textit{Philosophy of Science}, 12(2), 72-97.

\im Casta\~{n}eda, H.-N. (1970). On the semantics of the ought-to-do. \textit{Synthese}, 21, 449-468.

\im Cisewski, J., Kadane, J.~B., Schervish, M.~J., Seidenfeld, T., \& Stern, R. (2018). Standards for modest Bayesian credences. \textit{Philosophy of Science}, 85(1), 53-78.

\im de Finetti, B., \& Savage, L.~J. (1972). How to choose the initial probabilities. Translated and appearing as chapter 8 of de Finetti, B. (1972), {\em Probability, induction and statistics: The art of guessing}, New York: John Wiley.

\im de Finetti, B. (1974). \textit{Theory of probability} (Vol. 1). John Wiley, New York.

\im Devroye, L., Gy\"{o}rfi, L., \& Lugosi, G. (1996). \textit{A probabilistic theory of pattern recognition}. Springer.


\im Elga, A. (2016). Bayesian humility. \textit{Philosophy of Science}, 83(3), 305-323.

\im Freedman, D.~A. (1963). On the asymptotic behavior of Bayes' estimates in the discrete case. \textit{Annals of Mathematical Statistics}, 34(4), 1386-1403.

\im Friedman, M. (1979). Truth and confirmation. {\em The Journal of Philosophy}, 76(7), 361-382.

\im Gaifman, H., \& Snir, M. (1982). Probabilities over rich languages, testing and randomness. \textit{Journal of Symbolic Logic}, 47(3), 495-548.

\im Ghosal, S., \& van der Vaart, A. W. (2017). \textit{Fundamentals of nonparametric Bayesian inference}. Cambridge University Press.

\im Goldman, A.~I. (1979). What is justified belief? In G. S. Pappas (Ed.), \textit{Justification and knowledge: New studies in epistemology} (pp. 1-25). Dordrecht: Reidel. Reprinted in Goldman, A. I. (2012), \textit{Reliabilism and contemporary epistemology}, Oxford University Press, 29-49.

\im Gy\"{o}rfi, L., Kohler, M., Krzyzak, A., \& Walk, H. (2006). \textit{A distribution-free theory of nonparametric regression}. Springer.

\im Hacking, I. (1965/2016). \textit{Logic of statistical inference}. Cambridge University Press.

\im Horty, J. F., 2001, {\em Agency and Deontic Logic}, Oxford University Press.

\im Huttegger, S.~M. (2015). Bayesian convergence to the truth and the metaphysics of possible worlds. \textit{Philosophy of Science}, 82(4), 587-601.

\im Huttegger, S.~M. (2017). \textit{The probabilistic foundations of rational learning}. Cambridge University Press.

\im Huttegger, S.~M. (2022). Rethinking convergence to the truth. \textit{Journal of Philosophy}, 119(7), 380-403.

\im Jaynes, E. T. (1968). Prior probabilities. \textit{IEEE Transactions on Systems Science and Cybernetics}, 4(3), 227-241. 

\im Joyce, J.~M. (1998). A nonpragmatic vindication of probabilism. {\em Philosophy of Science}, 65(4), 575-603.

\im Joyce, J.~M. (2011). The development of subjective Bayesianism. In D. M. Gabbay, S. Hartmann, \& J. Woods (Eds.), \textit{Handbook of the history of logic: Inductive logic} (Vol. 10, pp. 415-475).

\im Kelly, K. T. (1996). {\em The logic of reliable inquiry}. Oxford University Press.

\im Konek, J., \& Levinstein, B. A. (2019). The foundations of epistemic decision theory. {\em Mind}, 128(509), 69-107.

\im Leplin, J. (2007). In defense of reliabilism. \textit{Philosophical Studies}, 134(1), 31-42.

\im Lewis, D. (1980). A subjectivist's guide to objective chance. In R. C. Jeffrey (Ed.), \textit{Studies in inductive logic and probability} (Vol. 2, pp. 263-293). Berkeley: University of California Press; reprinted in \textit{Philosophical papers, Vol. 2}, 1986.

\im Lin, H. (2022). Modes of convergence to the truth: Steps toward a better epistemology of induction. \textit{Review of Symbolic Logic}, 15(2), 277-310.

\im Lin, H. (2024). To be a frequentist or Bayesian? Five positions in a spectrum. \textit{Harvard Data Science Review}, 6(3), doi: 10.1162/99608f92.9a53b923

\im Lin, H. (2025). Convergence to the truth. In K. Sylvan, E. Sosa, J. Dancy, \& M. Steup (Eds.), \textit{The Blackwell companion to epistemology} (3rd ed.). Wiley Blackwell.


\im Mohri, M., Rostamizadeh, A., \& Talwalkar, A. (2018). \textit{Foundations of machine learning} (2nd edition). MIT Press.


\im Pettigrew, R. (2016). \textit{Accuracy and the laws of credence}. Oxford University Press.



\im Ramsey, F. P. (1926). Truth and probability. In Ramsey (1931). {\em The Foundations of Mathematics and other Logical Essays}, Ch. VII, pp. 156-198, edited by R.B. Braithwaite, London: Kegan, Paul, Trench, Trubner \& Co., New York: Harcourt, Brace and Company.

\im Reichenbach, H. (1938). \textit{Experience and prediction}. The University of Chicago Press.

\im Salmon, W. C. (1990). Rationality and objectivity in science or Tom Kuhn meets Tom Bayes. In C. W. Savage (Ed.), \textit{Scientific theories}, Minnesota Studies in the Philosophy of Science (pp. 175-205). Minneapolis: University of Minnesota Press.

\im Savage, L.~J. (1972). \textit{The foundations of statistics} (2nd ed.). Dover Publications.

\im Schroeder, M., 2011, ``Ought, Agents, and Actions'', {\em The Philosophical Review}, 120(1): 11-41.

\im Schulte, O. (1999). Means-ends epistemology. {\em The British Journal for the Philosophy of Science}, 50(1), 1-31.

\im Shalev-Shwartz, S., \& Ben-David, S. (2014). {\em Understanding machine learning: From theory to algorithms}. Cambridge University Press.

\im Shimony, A. (1970). Scientific inference. In R. G. Colodny (Ed.), \textit{The nature and function of scientific theories}, Pittsburgh Studies in the Philosophy of Science (Vol. 4, pp. 79-172). Pittsburgh: University of Pittsburgh Press.

\im Skyrms, B. (1980). \textit{Causal necessity}. New Haven: Yale University Press.


\im Titelbaum, M. G. (2013). {\em Quitting certainties: A Bayesian framework modeling degrees of belief}. Oxford University Press.

\im van Fraassen, B. C. (1989). \textit{Laws and symmetry}. Oxford University Press.

\im Wasserman, L. (1998). Asymptotic properties of nonparametric Bayesian procedures. In D. Dey, P. M\"{u}ller, \& D. Sinha (Eds.), \textit{Practical nonparametric and semiparametric Bayesian statistics}, Lecture Notes in Statistics (Vol. 133, pp. 293-304). Springer-Verlag New York.

\im Weatherson, B. (2015). For Bayesians, rational modesty requires imprecision. \textit{Ergo}, 2(20).

\im Wedgwood, R. (2006). The meaning of `ought'. {\em Oxford Studies in Metaethics}, 1: 127-160.

\im Wenmackers, S. \& Romeijn, J. W. (2016). New theory about old evidence: A framework for open-minded Bayesianism. {\em Synthese}, 193(4): 1225-1250.

\im Williamson J. (1999). Countable additivity and subjective probability. {\em The British Journal for the Philosophy of Science}, 50(3):401-416.

\im Williamson, J. (2010). \textit{In defense of objective Bayesianism}. Oxford University Press.
\end{description}

\end{document}